# Low Resistance Ohmic contact on epitaxial MOVPE grown β-Ga$_2$O$_3$ and β-(Al$_x$Ga$_{1-x}$)$_2$O$_3$ films

Fikadu Alema, Carl Peterson, Arkka Bhattacharyya, Saurav Roy, Sriram Krishnamoorthy and Andrei Osinsky

*Abstract*— We report on the realization of record low resistance Ohmic contacts to MOVPE-grown heavily Si-doped β-Ga$_2$O$_3$ and β-(Al$_x$Ga$_{1-x}$)$_2$O$_3$ epitaxial films. Transfer length measurement (TLM) patterns were fabricated on the heavily Si-doped homoepitaxial β-Ga$_2$O$_3$ films with electron concentration (n) ranging from 1.77 to 3.23×10$^{20}$ cm$^{-3}$. Record low specific contact resistance ($\rho_c$) and total contact resistance ($R_c$) of 1.62×10$^{-7}$ Ω.cm$^2$ and 0.023 Ω.mm were realized for β-Ga$_2$O$_3$: Si films with n >3×10$^{20}$ cm$^{-3}$. TLM structures were also fabricated on heavily Si doped coherently strained β-(Al$_x$Ga$_{1-x}$)$_2$O$_3$/β-Ga$_2$O$_3$ (x=12%, 17% and 22%) films. The film with 12% Al composition (n=1.23×10$^{20}$ cm$^{-3}$) showed $\rho_c$ of 5.85x10$^{-6}$ Ω.cm$^2$, but it increased to 2.19x10$^{-4}$ Ω.cm$^2$ for a layer with a 22% Al composition. Annealing the samples post metal deposition has generally led to a decrease in contact resistance, but for high Al content β-(Al$_x$, Ga$_{1-x}$)$_2$O$_3$, the contact resistance did not change significantly after the annealing process. The low contact resistance values measured in this work are very promising for the fabrication of high frequency power devices.

## I. INTRODUCTION

Beta gallium oxide (β-Ga$_2$O$_3$) has received significant attention in semiconductor research for applications in power electronic devices due to its fundamental properties, including a high breakdown field of ~ 8 MV/cm[1], availability of high quality melt grown native substrates [2], and controllable donor doping [3, 4]. The availability of high quality Ga$_2$O$_3$ substrates has resulted in the rapid development of high quality β-Ga$_2$O$_3$ epilayers and devices [5-8]. Excellent device results, such as breakdown voltages and critical fields exceeding 2.5 kV and 3 MV/cm, have been reported on numerous lateral and vertical β-Ga$_2$O$_3$ power devices [6, 9-12]. However, as a gateway to the external world, reliable low resistance Ohmic contacts are critical for the efficient performance of any device. High contact resistance at the metal/β-Ga$_2$O$_3$ junction leads to slower device switching speeds and device failure due to local contact heating from a significant voltage drop at the junction during device operation[13, 14].

F. Alema and C. Peterson contributed equally to this work.
This work was supported by Air Force Research Laboratory (AFRL)/RQKMA(FA8650-17-F-5418) and II-VI foundation Block Gift Program 2021-2022. C. Peterson would like to acknowledge Esmat Farzana for help with TLM fabrication process.
F. Alema and A. Osinsky are with Agnitron Technology Incorporated, Chanhassen, MN 55317, USA.
C. Peterson, S. Roy, and S. Krishnamoorthy are with Materials Department, University of California, Santa Barbara, CA 93106, USA.
A. Bhattacharyya is with the Department of Electrical and Computer Engineering, University of Utah, Salt Lake City, Utah, USA 84112.
* Corresponding author e-mail: Fikadu. Alema@agnitron.com, carlpeterson@ucsb.edu

Various techniques, including ion implantation [15], spin-on-glass [16], and regrowth methods [5, 17], have been employed to reduce contact resistance at the source/drain (S/D) ohmic contacts in Ga$_2$O$_3$ MOSFETs and MESFETs. Both ion implantation and spin-on-glass methods expose the material to a high annealing temperature (~900-1200 °C) and potentially deteriorate the quality of the device's active region [15, 16]. However, regrowth process is typically performed at a much lower substrate temperature (~600 °C) and a low contact resistance can be achieved without affecting the material quality. Recently, MOVPE based regrowth process has been applied to fabrication of various FETs, realizing a metal/n$^+$-Ga$_2$O$_3$ contact resistance as low as 0.08 Ω. mm and $\rho_c$ of ~8×10$^{-7}$ Ω. cm$^2$ [5]. However, a systematic study to heavily dope β-Ga$_2$O$_3$ epitaxial layers using MOVPE and achieve low contact resistance is still lacking. In this work, we report on the demonstration of record low resistance Ohmic contacts on heavily Si doped epitaxial β-Ga$_2$O$_3$ and pseudomorphic Si doped β-(Al$_x$Ga$_{1-x}$)$_2$O$_3$ epilayers with varying Al composition.

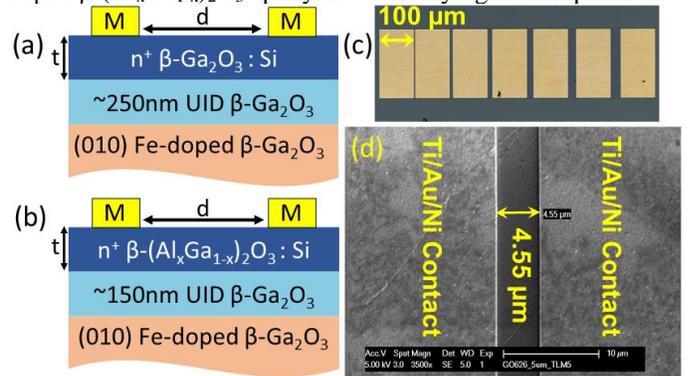

Fig. 1. Schematic cross-sectional view of the heavily Si doped (a) β-Ga$_2$O$_3$ and (b) β-(Al$_x$Ga$_{1-x}$)$_2$O$_3$/β-Ga$_2$O$_3$ heterostructure with Ti/Au/Ni metal stack for Ohmic contact. (c) Optical microscope image showing the TLM pattern (d) SEM image verifying the gap spacing, d.

## II. MATERIALS GROWTH AND DEVICE FABRICATION

Three heavily Si doped homoepitaxial β-Ga$_2$O$_3$ films (samples A, B, and C, see Table I) and three fully strained heavily Si doped β-(Al$_x$Ga$_{1-x}$)$_2$O$_3$/β-Ga$_2$O$_3$ heterostructures with Al composition (*x*) of 12% (Sample D), 17% (sample E), and 22% (sample F) were grown using Agnitron Technology's Agilis 100 MOVPE reactor on (010) β-Ga$_2$O$_3$:Fe substrates. The doped β-Ga$_2$O$_3$ and β-(Al$_x$Ga$_{1-x}$)$_2$O$_3$ layers were grown at ~600 °C on an ~250 nm and ~150 nm thick unintentionally doped (UID) Ga$_2$O$_3$ buffer layers (Fig. 1), respectively, with TEGa, TEAl, O$_2$, and silane (SiH$_4$) as precursors, and Ar as carrier gas. The layer thickness and doping concentration for the heavily Si doped β-Ga$_2$O$_3$ and β-(Al$_x$Ga$_{1-x}$)$_2$O$_3$ films are shown in Table I. The purity of the phase, Al composition (*x*), and layer thickness for the β-(Al$_x$Ga$_{1-x}$)$_2$O$_3$ layers were determined using HRXRD [18]. Hall effect measurements were performed on each of the samples to determine their electron concentration (n), mobility ($\mu_e$), and sheet resistance ($R_{sh}$) as presented in Table I. The films were then processed into linear transmission line model (TLM) test structures to characterize their contact resistance. TLM structures were mesa isolated to the β-Ga$_2$O$_3$ substrate using BCl$_3$ chemistry-based reactive ion etching (RIE) process. A 20 nm/150 nm/50 nm Ti/Au/Ni metal stack was deposited via e-beam evaporation and was annealed at 470 °C in N$_2$ for 1min. Four probe current – voltage (I-V) measurements were performed on the TLM structures to obtain the specific contact resistance ($\rho_c$) of the ohmic contact.

## III. RESULTS AND DISCUSSIONS

The TLM measurements were performed at room temperature before and after rapid annealing of the contacts. Fig. 2(a) and 2(b) show the total resistance between electrodes as a function of electrode spacing, d, for representative β-Ga$_2$O$_3$ epilayer (sample C, n = 3.23×10$^{20}$ cm$^{-3}$) and β-(Al$_{0.12}$Ga$_{0.88}$)$_2$O$_3$/β-Ga$_2$O$_3$ heterostructure (sample D, n = 1.23×10$^{20}$ cm$^{-3}$) samples after contact annealing. The output resistance has shown strong linear dependance on electrode spacing for all the samples, and the TLM data was fitted to extract sheet resistance ($R_{sh}$), specific contact resistance ($\rho_c$), total contact resistance ($R_c$), and transfer length ($L_t$) for the respective samples. For the β-Ga$_2$O$_3$ and β-(Al$_{0.12}$Ga$_{0.88}$)$_2$O$_3$ layers $\rho_c$ of ~10$^{-7}$ Ω.cm$^2$ and ~6×10$^{-6}$ Ω.cm$^2$, respectively, were obtained.

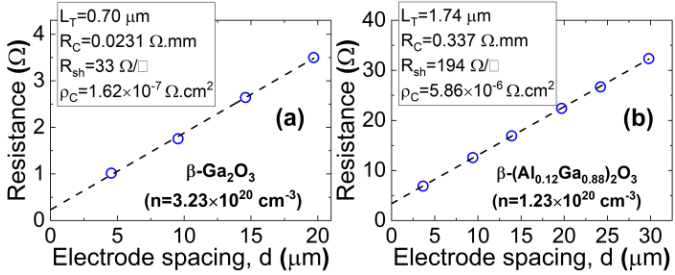

Fig. 2. Total resistance between electrodes versus electrode spacing for heavily Si doped MOVPE grown: (a) β-Ga$_2$O$_3$ (n = 3.23×10$^{20}$ cm$^{-3}$) and (b) β-(Al$_{0.12}$Ga$_{0.88}$)$_2$O$_3$ (n = 1.23×10$^{20}$ cm$^{-3}$) measured after metallization contact anneal. The circular symbols are measured data and the broken lines are linear fitting to extract $R_{sh}$, $R_c$, $\rho_c$ and $L_t$ parameters.

Table I. Electron concentration (n), Hall electron mobility ($\mu_e$), and sheet resistance ($R_{sh-Hall}$) of the Ga$_2$O$_3$ and β-(Al$_x$Ga$_{1-x}$)$_2$O$_3$ layers measured by Hall effect measurements. The layers thickness (t) and sheet resistance measured by TLM method ($R_{sh-TLM}$) is also presented.

| Samples | t (nm) | n (cm$^{-3}$) | $\mu_e$ (cm$^2$ V$^{-1}$s$^{-1}$) | $R_{sh-Hall}$ (Ω/□) | $R_{sh-TLM}$ (Ω/□) |
|---|---|---|---|---|---|
| A | 145 | 1.77×10$^{20}$ | 25.2 | 97 | 107 |
| B | 65 | 2.51×10$^{20}$ | 53.1 | 72 | 87 |
| C | 170 | 3.23×10$^{20}$ | 38.2 | 30 | 33 |
| D | 75 | 1.23×10$^{20}$ | 31.1 | 216 | 194 |
| E | 110 | 1.22×10$^{20}$ | 16.2 | 284 | 290 |
| F | 84 | 5.49×10$^{19}$ | 24.9 | 608 | 514 |

The sheet resistance ($R_{sh}$) extracted from the TLM after contact annealing is compared with the values measured from Hall effect measurements (Table I). The $R_{sh}$ values obtained from the two methods are very comparable, showing the consistency of the results obtained for each film. Fig 3. shows the specific contact resistance ($\rho_c$) and total contact resistance ($R_c$) measured post contact annealing as a function of electron concentration for Ga$_2$O$_3$:Si. Both $\rho_c$ and $R_c$ decreased with the increase in doping concentration. All the Ga$_2$O$_3$ epitaxial layers studied in this work (n>1.7×10$^{20}$ cm$^{-3}$) demonstrated $\rho_c$ of < 1.66×10$^{-6}$ Ω.cm$^2$ (the lowest being 1.62×10$^{-7}$ Ω.cm$^2$), much lower than all the specific contact resistance values reported in the literature so far [5, 15, 19-21]. The lowest specific contact resistance ($\rho_c$) of 1.62×10$^{-7}$ Ω.cm$^2$ obtained in this work is lower than the best specific constant resistance reported by the MOVPE regrowth method by ~4× ref. [5] and ion implantation method by 23× ref. [15]. The $R_c$ values were also lower than 0.05 Ω.mm for Ga$_2$O$_3$ samples with n >2.5×10$^{20}$ cm$^{-3}$. Both $\rho_c$ and Rc values obtained in this work are the lowest among UWBG materials, suggesting that MOVPE- regrown contacts could play a significant role in the future development of efficient and fast switching RF devices.

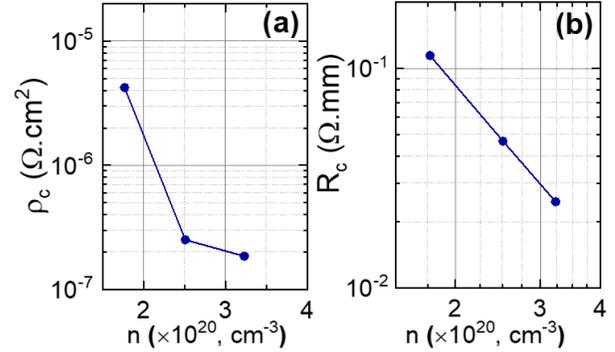

Fig. 3. (a) $\rho_c$ and (b) $R_c$ as a function of carrier concentration for heavily Si doped β-Ga$_2$O$_3$ measured after metallization contact anneal.

Fig. 4(a) and 4(b) presents the specific contact resistance ($\rho_c$) and total contact resistance ($R_c$) measured post contact annealing as a function of Al composition for β-(Al$_x$Ga$_{1-x}$)$_2$O$_3$/β-Ga$_2$O$_3$ heterostructures. The left panel (y-axis) for both figures show the measured electron concentration (n). The plots also include the $\rho_c$, $R_c$, and n for pure Ga$_2$O$_3$ to indicate β-(Al$_x$Ga$_{1-x}$)$_2$O$_3$ with x=0. With the increase in the Al composition of the β-(Al$_x$Ga$_{1-x}$)$_2$O$_3$ from 0 to 22%, the n decreased, and the $\rho_c$ and $R_c$ increased. When the Al composition in the β-(Al$_x$Ga$_{1-x}$)$_2$O$_3$ increased from 0 (i.e., pure Ga$_2$O$_3$) to 12% (i.e., β-(Al$_{0.12}$Ga$_{0.88}$)$_2$O$_3$) the $\rho_c$ increased by more than an order of magnitude (from 1.62×10$^{-7}$ to 5.86×10$^{-6}$ Ω.cm$^2$) but was still comparable to the best $\rho_c$ reported for Si implanted Ga$_2$O$_3$ [15]. For the β-(Al$_x$Ga$_{1-x}$)$_2$O$_3$ with Al composition of 22%, the specific contact resistance obtained was 2.19×10$^{-4}$ Ω.cm$^2$. Similarly, the total contact resistance increased with the increase in Al composition.

It is a common practice to perform rapid thermal annealing following metallization at a temperature ranging from 400 to 500 °C to improve the contact resistance at the metal/n$^+$-Ga$_2$O$_3$ junction [15, 19, 22]. In this work, as indicated above, we measured the contact resistance of each of the samples before and after annealing the Ti/Au/Ni metal stack contacts in N$_2$ at 470 °C to study the effect annealing on the β-(Al$_x$Ga$_{1-x}$)$_2$O$_3$ layers. Table II compares the $\rho_c$ for the Ga$_2$O$_3$ and β-(Al$_x$Ga$_{1-x}$)$_2$O$_3$ samples. For the Ga$_2$O$_3$ films, the $\rho_c$ decreased following the annealing process, showing the expected improvement in contact resistance. But, for β-(Al$_x$Ga$_{1-x}$)$_2$O$_3$ layers, a significant decrease in $\rho_c$ after annealing was observed only for a film with lower Al composition (12%). For higher Al composition β-(Al$_x$Ga$_{1-x}$)$_2$O$_3$, the $\rho_c$ stayed relatively the same or increased following the annealing process (See Table II). This is likely due to the differences in interfacial reaction and interface chemical composition of Ti/AlGaO as compared to Ti/Ga$_2$O$_3$ annealed interfaces[14].

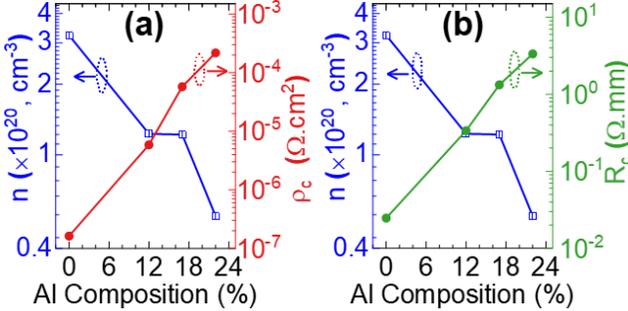

Fig. 4. (a) $\rho_c$ and (b) R$_c$ as a function of Al composition measured after metallization contact anneal. The left panel on both figures shows electron concentration (n) dependance on Al composition. The values at x=0 is for pure Ga$_2$O$_3$.

Table II. Compares $\rho_c$ values measured before and after contact annealing. R$_c$ measured after the contact annealing is shown.

| Samples | $\rho_c$ (Ω. cm$^2$) | | R$_c$ (Ω. mm) |
| --- | --- | --- | --- |
| | [Pre-anneal] | [Post-anneal] | [Post-anneal] |
| A | 4.23×10$^{-6}$ | 1.66×10$^{-6}$ | 0.110 |
| B | 1.68×10$^{-6}$ | 2.51×10$^{-7}$ | 0.047 |
| C | 1.12×10$^{-6}$ | 1.62×10$^{-7}$ | 0.023 |
| D | 1.30×10$^{-5}$ | 5.86×10$^{-6}$ | 0.340 |
| E | 3.56×10$^{-5}$ | 5.77×10$^{-5}$ | 1.270 |
| F | 3.96×10$^{-4}$ | 2.19×10$^{-4}$ | 3.350 |

A heavily Ge doped MOVPE grown β-Ga$_2$O$_3$ film (GeH$_4$/N$_2$ as Ge source) with n=2.6×10$^{20}$ cm$^{-3}$ and μ$_e$ ~38 cm$^2$/Vs [18, 23] was also fabricated into TLM and $\rho_c$ and R$_c$ values of 2.1×10$^{-6}$ (Ω. cm$^2$) and 0.06 (Ω. mm) were obtained. Although less desirable as a dopant for Ga$_2$O$_3$ due to its strong process dependence and severe memory effect, Ge can still be used to obtain low resistance Ohmic contacts for FETs [6, 10, 23].

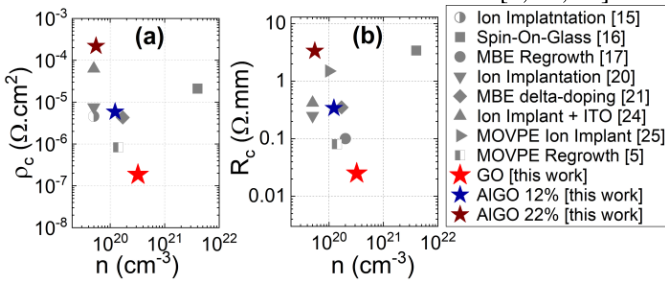

Fig. 5. $\rho_c$ (a) and R$_c$ (b) from this work as compared to the best reported results from various methods [5, 15-17, 20, 21, 24, 25]. Results measured for β-(Al$_x$Ga$_{1-x}$)$_2$O$_3$ films with Al composition of 12% and 22% are also included.

Fig. 5(a) and 5(b) benchmarks our specific contact resistance ($\rho_c$) and total contact resistance (R$_c$) values with the existing literature reports. The comparison shows that the obtained $\rho_c$ and R$_c$ values for the Ga$_2$O$_3$ is the lowest of all the values reported. Even for β-(Al$_{0.12}$Ga$_{0.88}$)$_2$O$_3$, the obtained $\rho_c$ and R$_c$ values are comparable to those reported for pure Ga$_2$O$_3$ using ion implantation method [20]. Thus, utilizing low temperature MOVPE epitaxy, heavily doped Ga$_2$O$_3$ and β-(Al$_x$Ga$_{1-x}$)$_2$O$_3$ epitaxial films can be grown to realize a low metal/semiconductor contact resistance. Such a result is very encouraging for high frequency devices where low parasitic resistance is critical.

## IV. Conclusion

We successfully demonstrated record low resistance Ohmic contacts to MOVPE-grown heavily Si doped β-Ga$_2$O$_3$ and β-(Al$_x$Ga$_{1-x}$)$_2$O$_3$ epitaxial films. For β-Ga$_2$O$_3$: Si with an electron concentration of 3.23×10$^{20}$ cm$^{-3}$, $\rho_c$ and R$_c$ values of 1.62x10$^{-7}$ Ω.cm$^2$ and 0.023 Ω.mm were obtained. For β-(Al$_x$Ga$_{1-x}$)$_2$O$_3$, the electron concentration was found to decrease with the increase in Al composition, and thus led to an increase in $\rho_c$ and R$_c$. The record low metal/semiconductor contact resistance measured both for β-Ga$_2$O$_3$ and β-(Al$_x$Ga$_{1-x}$)$_2$O$_3$ in this work will have significant impact in advancing the performance of RF devices, where low parasitic resistances is paramount.

## References

[1] M. Higashiwaki, K. Sasaki, A. Kuramata, T. Masui, and S. Yamakoshi, "Gallium oxide (Ga2O3) metal-semiconductor field-effect transistors on single-crystal β-Ga2O3 (010) substrates," *Applied Physics Letters,* vol. 100, no. 1, p. 013504, January.2012.10.1063/1.3674287

[2] A. Kuramata, K. Koshi, S. Watanabe, Y. Yamaoka, T. Masui, and S. Yamakoshi, "High-quality β-Ga2O3 single crystals grown by edge-defined film-fed growth," *Japanese Journal of Applied Physics,* vol. 55, no. 12, p. 1202A2, November.2016.10.7567/jjap.55.1202a2

[3] F. Alema, A. Osinsky, N. Orishchin, N. Valente, Y. Zhang, A. Mauze, and J. S. Speck, "Low 1014 1/cm3 free carrier concentration in MOCVD grown epitaxial β-Ga2O3 " *APL Materials,* vol. 8, pp. 021110-9 February.2020.10.1063/1.5132752

[4] A. T. Neal, S. Mou, S. Rafique, H. Zhao, E. Ahmadi, J. S. Speck, K. T. Stevens, J. D. Blevins, D. B. Thomson, N. Moser, K. D. Chabak, and G. H. Jessen, "Donors and deep acceptors in β-Ga2O3," *Applied Physics Letters,* vol. 113, no. 6, p. 062101, August.2018.10.1063/1.5034474

[5] A. Bhattacharyya, S. Roy, P. Ranga, D. Shoemaker, Y. Song, J. S. Lundh, S. Choi, and S. Krishnamoorthy, "130 mA mm−1 β-Ga2O3 metal semiconductor field effect transistor with low-temperature metalorganic vapor phase epitaxy-regrown ohmic contacts," *Applied Physics Express,* vol. 14, no. 7, p. 076502, 2021/06/22.2021.10.35848/1882-0786/ac07ef

[6] A. Bhattacharyya, S. Sharma, F. Alema, p. ranga, S. Roy, C. Peterson, G. Seryogin, A. Osinsky, U. Singisetti, and S. Krishnamoorthy, "4.4 kV β-Ga2O3 MESFETs with Power Figure of Merit exceeding 100 MW/cm2," *Applied Physics Express,* vol. 15, p. 061001, February 2022.10.35848/1882-0786/ac6729

[7] G. Seryogin, F. Alema, N. Valente, H. Fu, E. Steinbrunner, A. T. Neal, S. Mou, A. Fine, and A. Osinsky, "MOCVD growth of high purity Ga2O3 epitaxial films using trimethylgallium precursor," *Applied Physics Letters,* vol. 117, no. 26, p. 262101, 2020.10.1063/5.0031484

[8] F. Alema, Y. Zhang, A. Osinsky, N. Valente, A. Mauze, T. Itoh, and J. S. Speck, "Low temperature electron mobility exceeding 104 cm2/V s in MOCVD grown β-Ga2O3," *APL Materials,* vol. 7, no. 12, pp. 121110:1-6, December.2019.10.1063/1.5132954

[9] A. J. Green, K. D. Chabak, E. R. Heller, R. C. Fitch, M. Baldini, A. Fiedler, K. Irmscher, G. Wagner, Z. Galazka, S. E. Tetlak, A. Crespo, K. Leedy, and G. H. Jessen, "3.8-MV/cm Breakdown Strength of MOVPE-Grown Sn-Doped b-Ga2O3 MOSFETs," *IEEE Electron Device Letters,* vol. 37, no. 7, p. 902, July.2016.10.1109/led.2016.2568139

[10] A. Bhattacharyya, P. Ranga, S. Roy, C. Peterson, F. Alema, G. Seryogin, A. Osinsky, and S. Krishnamoorthy, "Multi-kV Class β-Ga2O3 MESFETs With a Lateral Figure of Merit Up to 355 MW/cm²," *IEEE Electron Device Letters,* vol. 42, no. 9, pp. 1272-1275, July.2021.10.1109/LED.2021.3100802

[11] Z. Xia, H. Chandrasekar, W. Moore, C. Wang, A. J. Lee, J. McGlone, N. K. Kalarickal, A. Arehart, S. Ringel, F. Yang, and S.

Rajan, "Metal/BaTiO3/β-Ga2O3 dielectric heterojunction diode with 5.7 MV/cm breakdown field," *Applied Physics Letters,* vol. 115, no. 25, p. 252104, December.2019.10.1063/1.5130669

[12] S. Sharma, K. Zeng, S. Saha, and U. Singisetti, "Field-Plated Lateral Ga2O3 MOSFETs With Polymer Passivation and 8.03 kV Breakdown Voltage," *IEEE Electron Device Letters,* vol. 41, no. 6, pp. 836-839, 2020.10.1109/led.2020.2991146

[13] S. J. Pearton, J. Yang, P. H. Cary, F. Ren, J. Kim, M. J. Tadjer, and M. A. Mastro, "A review of Ga2O3materials, processing, and devices," *Applied Physics Reviews,* vol. 5, no. 1, 2018.10.1063/1.5006941

[14] M.-H. Lee and R. L. Peterson, "Process and characterization of ohmic contacts for beta-phase gallium oxide," *Journal of Materials Research,* vol. 36, no. 23, pp. 4771-4789, 2021.10.1557/s43578-021-00334-y

[15] K. Sasaki, M. Higashiwaki, A. Kuramata, T. Masui, and S. Yamakoshi, "Si-Ion Implantation Doping in β-Ga2O3and Its Application to Fabrication of Low-Resistance Ohmic Contacts," *Applied Physics Express,* vol. 6, no. 8, 2013.10.7567/apex.6.086502

[16] K. Zeng, J. S. Wallace, C. Heimburger, K. Sasaki, A. Kuramata, T. Masui, J. A. Gardella, and U. Singisetti, "Ga2O3MOSFETs Using Spin-On-Glass Source/Drain Doping Technology," *IEEE Electron Device Letters,* vol. 38, no. 4, pp. 513-516, 2017.10.1109/led.2017.2675544

[17] Z. Xia, C. Joishi, S. Krishnamoorthy, S. Bajaj, Y. Zhang, M. Brenner, S. Lodha, and S. Rajan, "Delta Doped β-Ga2O3 Field-Effect Transistors With Regrown Ohmic Contacts," *IEEE Electron Device Letters,* vol. 39, no. 4, pp. 568-571, April.2018.10.1109/LED.2018.2805785

[18] F. Alema, T. Itoh, J. S. Speck, and A. Osinsky, "Highly conductive epitaxial β-Ga2O3 and (AlxGa1-x)2O3 films by MOCVD," in *APEX,* ed, 2022.

[19] M. Higashiwaki, K. Sasaki, T. Kamimura, M. Hoi Wong, D. Krishnamurthy, A. Kuramata, T. Masui, and S. Yamakoshi, "Depletion-mode Ga2O3 metal-oxide-semiconductor field-effect transistors on β-Ga2O3 (010) substrates and temperature dependence of their device characteristics," *Applied Physics Letters,* vol. 103, no. 12, 2013.10.1063/1.4821858

[20] M. H. Wong, Y. Nakata, A. Kuramata, S. Yamakoshi, and M. Higashiwaki, "Enhancement-mode Ga2O3 MOSFETs with Si-ion-implanted source and drain," *Applied Physics Express,* vol. 10, no. 4, 2017.10.7567/apex.10.041101

[21] S. Krishnamoorthy, Z. Xia, S. Bajaj, M. Brenner, and S. Rajan, "Delta-doped β-gallium oxide field-effect transistor," *Applied Physics Express,* vol. 10, no. 5, 2017.10.7567/apex.10.051102

[22] M.-H. Lee and R. L. Peterson, "Annealing Induced Interfacial Evolution of Titanium/Gold Metallization on Unintentionally Dopedβ-Ga2O3," *ECS Journal of Solid State Science and Technology,* vol. 8, no. 7, pp. Q3176-Q3179, 2019.10.1149/2.0321907jss

[23] F. Alema, G. Seryogin, A. Osinsky, and A. Osinsky, "Ge doping of β-Ga2O3 by MOCVD," *APL Materials,* vol. 9, no. 9, pp. 091102:1-11 September.2021.10.1063/5.0059657

[24] P. H. Carey, J. Yang, F. Ren, D. C. Hays, S. J. Pearton, A. Kuramata, and I. I. Kravchenko, "Improvement of Ohmic contacts on Ga2O3through use of ITO-interlayers," *Journal of Vacuum Science & Technology B, Nanotechnology and Microelectronics: Materials, Processing, Measurement, and Phenomena,* vol. 35, no. 6, 2017.10.1116/1.4995816

[25] K. J. Liddy, A. J. Green, N. S. Hendricks, E. R. Heller, N. A. Moser, K. D. Leedy, A. Popp, M. T. Lindquist, S. E. Tetlak, G. Wagner, K. D. Chabak, and G. H. Jessen, "Thin channel β-Ga2O3 MOSFETS with self-aligned refractory metal gates," *Applied Physics Express,* vol. 12, no. 12, p. 126501, 2019/10/29.2019.10.7567/1882-0786/ab4d1c